# Augmented reality as a tool for open science platform by research collaboration in virtual teams

Mariya P. Shyshkina[0000-0001-5569-2700] and Maiia V. Marienko[0000-0002-8087-962X]

Institute of Information Technologies and Learning Tools of the NAES of Ukraine,
9, M. Berlynskoho Str., Kyiv, 04060, Ukraine
`{shyshkina, popel}@iitlt.gov.ua`

**Abstract.** The provision of open science is defined as a general policy aimed at overcoming the barriers that hinder the implementation of the European Research Area (ERA). An open science foundation seeks to capture all the elements needed for the functioning of ERA: research data, scientific instruments, ICT services (connections, calculations, platforms, and specific studies such as portals). Managing shared resources for the community of scholars maximizes the benefits to society. In the field of digital infrastructure, this has already demonstrated great benefits. It is expected that applying this principle to an open science process will improve management by funding organizations in collaboration with stakeholders through mechanisms such as public consultation. This will increase the perception of joint ownership of the infrastructure. It will also create clear and non-discriminatory access rules, along with a sense of joint ownership that stimulates a higher level of participation, collaboration and social reciprocity. The article deals with the concept of open science. The concept of the European cloud of open science and its structure are presented. According to the study, it has been shown that the structure of the cloud of open science includes an augmented reality as an open-science platform. An example of the practical application of this tool is the general description of MaxWhere, developed by Hungarian scientists, and is a platform of aggregates of individual 3D spaces.

## 1 Introduction

In order for researchers to be able to focus on their work, newly developed electronic computing resources and cloud services should not only offer the functions necessary to solve the problems of large data, but also work smoothly and intuitively, without emphasizing the technical details of the cloud-based environments Thus, today's demands of the research and education community require a holistic approach in the development of the next generation of intelligent networks, which should work in concert with the components of distributed application.

## 1.1 The problem statement

Calculations that are traditionally used to store and process large amounts of data remain difficult to use, both in terms of programming and in terms of data management. This is especially emphasized by the latest trends in modern research, which are becoming more and more manageable and associated with big data [10]. The latter require the processing of a huge amount of distributed computing in an easy way. Most of the current high-tech data tasks can easily be rolled into a list of independent tasks that can be handled in parallel (for example, using a cloud platform and do not require additional software), while the problem of distributed computing, storage and fast data remains unresolved.

In order to focus on their research, researchers need to be able to analyze and process data specific to the program intuitively. Users do not need to understand the core cloud infrastructure software blocks that need to deal with distributed computing, storage, and interconnection issues. The examples that cover these problems can be found in virtually all branches of science, such as bioinformatics, geological science, high-quality streaming video and real-time processing, or the design work of a large group of scientists geographically distant from one another [4].

Cloud computing in all of its available models, such as IaaS, PaaS and SaaS [5], plays an important role in this attempt to facilitate collaborative research by not exploring and managing the details of the underlying infrastructure in order to be able to use it for joint data processing. By providing abstraction of resources and simple automation tools, modern cloud platforms simplify most routing tasks such as installation, maintenance, backup, security, and more. Thus, cloud applications have become an important tool for modern researchers. Moreover, today, they are, as a rule, the best way to solve the problem of big data [4].

To solve research-related problems, modern science needs support from computing infrastructures, so many European and national initiatives deal with distributed, networked and cloud-based infrastructures. One of them is the Helix-Nebula project, the European Network Infrastructure (EGI), the European Open Cloud of Science (EOSC-hub). Due to the high demand for research applications, similar services related to data storage, for the processing of a huge amount of data are increasing interest from the scientific community. It is expected that these services will provide both productivity and features that allow more flexible and cost-effective use of such services. Easy multi-platform data access, long-term storage, performance support, and cost of data access are elements that can be differentiated into one system. In order to meet the needs of the scientific community regarding infrastructure in Poland, several national projects were also launched. The results of the PL-Grid family of projects provide a computing infrastructure for large-scale simulations and calculations at high-performance computing clusters supported by domain-based services, solutions and environments. Pioneer's infrastructure serves high-bandwidth optical networks that connect the main computer centers used in the infrastructure of PL-Grid. Since the scientific data obtained through simulation, sensors or devices used by scientific applications should be stored for further research in appropriate repositories, such services are in high demand [7]. Some requirements put forward by users relate to aspects of service quality and its proper level [8].

### 1.2 Analysis of recent research and publications

For Ukrainian science, issues relating to the European cloud of open science are new and little studied. However, certain work is already being met and scientists are actively interested in the issues. Olekcey O. Petrenko [9] investigated the changes taking place in service-oriented architectures in connection with the transfer of applied applications in the cloud environment, in particular, to the European cloud of open science.

Valerii Yu. Bykov [2] investigated the scientific and methodological basis for the creation and development of a cloud-based environment in the context of open scientific priorities and the formation of the European Research Area (ERA). Their work outlines the conceptual and terminological justification of cloud computing, as well as the main features of such a medium. Ukrainian scientists describe the main methodological principles of designing and developing the environment, on the example of the principles of open science, open education, as well as the specific principles inherent in cloud-based systems.

### 1.3 The purpose of the article

On the basis of analysis of the structure of an open science platform, it is shown that complemented reality serves as its tool and on a separate software product to determine its practical value in scientific research.

## 2 Theoretical background

Scientists around the world are increasingly using cloud-based technologies to perform computational tasks. Cloud resources can be distributed on demand, scaled according to different usage patterns, and reduced costs for individual groups of scientists to support their own infrastructure.

Olekcey O. Petrenko in [9, p. 13-14] notes that service-oriented approach that is based on the present-day largest European project for the creation of the European Open Science Cloud for Research (EOS), which began in 2017 and which motivates research into the technology of hosting many SOA applications in the cloud, which will soon serve 1,7 million scientists and 80 million professionals from various fields of science and technology.

Major research infrastructures are planned on an EU-wide scale in the context of the ESFRI roadmap, aimed at providing scientists with the appropriate tools for research. More and more demands on data volumes and computing power are put forward.

Projects such as Indigo-Datacloud, EGI, European Cloud Science, HelixNebula, are considering the introduction of cloud services for the European academic community [1].

Indigo-DataCloud develops intermediate software for implementing a variety of cloud-based services, from authentication, workload and data management, and collects a catalog of cloud services. The project just released the second software, ElectricIndigo.

The Indigo project is primarily aimed at bridging the gap between cloud-developers and the services provided by existing cloud service providers, instead of providing their own cloud-based services.

EGI coordinates a unified cloud, originally based on OCCI and CDMI, as web services interfaces to access OpenNebula and OpenStack cluster resources or public service providers. This approach is to provide an additional level of abstraction over the resources provided by national energy conservation programs and remain separate and independent of each other.

HelixNebula explores how best to use commercial cloud service providers in the purchase of cloud infrastructure for research and education. This approach is to create a private-government partnership for the purchase of hybrid clouds.

The third phase of the prototype, which involves three contract consortia, has recently begun. The European Commission promotes the European cloud of open science as a common basis for supporting open science and research, covering a wide range of issues ranging from technical, accessible and managerial to building infrastructure. Many of these projects are funded by research or meet the needs of specific communities, such as providing prototype or pilot-level services to a limited group of users, with limited resources, as well as groups within the EGI Federated Cloud Initiative. Moving from the prototype stage to the production stage, offering large volumes of resources for a large community, is a challenge in terms of efforts and resources. Creating a well-equipped and supported platform for cloud computing requires a significant investment of large commercial cloud providers or public organizations that decide to invest in creating a real cloud infrastructure for science. One of the possible alternatives to a central approach to large-scale financing is the federative approach, where the infrastructure is built up from the bottom up, combining medium / large objects into large ones, to reach the appropriate scale [1].

Within the framework of the European Commission's strategy for creating a single digital market, the European Commission officially launched the European Open Educational Initiative (EOSC) in April 2016. EOSC promotes not only scientific excellence and data reuse, but also job creation and competitiveness in Europe, as well as contributing to pan-European cost efficiencies in scientific infrastructures by promoting unprecedented scale.

The experts outlined the basic principles of the cloud of open science [3]:

1. EOSC needs to integrate with other electronic infrastructures and initiatives in the world by introducing a light, interconnected system of services and data that fits the federal model.
2. The term "open" refers to the availability of services and data in accordance with the appropriate non-discriminatory policy ("not all data and tools may be open", and "free data and services do not exist").
3. The EOSC should include all academic disciplines.

4. The term "cloud" should not relate to ICT infrastructure, but to universal access to data, software, standards, expertise and policy frameworks for science and innovation-driven data.

The general view of most relevant stakeholders for the European cloud of open science lies in the fact that this cloud should [9]:

- to be a system of services provided by different suppliers;
- relying on existing electronic infrastructures, so developer efforts should focus on the integration / interoperability of cloud services;
- continuously develop and integrate new services and tools as soon as they become available, freely distributed to users;
- to take into account the needs of users as a leading motive for the development of the European cloud of open science.

In the vision of experts, EOSC will be an accessible infrastructure for modern research and innovation that employs the Internet of accessible data and interoperability and reusable services. It should be based on standards, best practices and infrastructures supplemented by adequate human experience. The fair principles should be maintained, and particular attention should be paid to the reuse of open and confidential data. Data should be with a multitude of elements (standard, tools, protocols) that provide the possibility and ease of reuse. In addition, there is a need to implement a science data processing profession to ensure professional data management and long-term management. In Europe, European research infrastructures specializing in the domain, and cross-sectoral ICT electronic infrastructures as well as other disciplinary and interdisciplinary collaborations and services have already been established. They can be considered the basis for EOSC. However, the implementation of ambitions to increase unimpeded access, reliable reuse of data and other digital research objects, as well as cooperation between different services and infrastructures (which guarantees non-discriminatory access and reuse of data both to the public and to the public and private sector), requires further improvement of this landscape in order to transform the ever-increasing amount of data on knowledge as a renewable, sustainable ground for innovation in turn to meet the global needs. EOSC is an instrument defined by the European Commission to facilitate such development towards the implementation of the Open Science. This idea highlights the strong link between ERA implementation through Open Science, Open Science and EOSC. In this context, the High-Level Expert Group, developed by the European Commission, reported on the list of key trends of Open Science that should be taken into account in the EOSC project. They cover several aspects, such as new ways of scientific communication (for example, programs, software conveyors and data itself), new incentives for promoting data dissemination and sharing of tools, facilitating the formation of data processing professionals, interdisciplinary collaboration, support for innovative SMEs, the creation of ecosystems, methodologies and tools for the reproduction of current published research, etc. [3].

## 3     Research methodology

The ERA was endorsed by the European Council in 2000 as a way of building a single, open-world research area based on a domestic market in which researchers, scientific knowledge and technology circulate freely and through which the European Union and its members strengthen their scientific and technological bases, their competitiveness and the ability to collectively address the challenges of today [3].

According to Olekcey O. Petrenko, EOSC is an interdisciplinary environment for research, innovation and educational goals [9, p. 59].

According to the first report of the High-Level Expert Group on the EOSC appointed by the European Commission, EOSC was identified as an open source support environment for accelerating the transition to more effective open science and open innovation in digital the single market by removing technical, legislative and human barriers to reuse data and research tools. Indeed, the term "cloud" was interpreted as a metaphor that helps convey the idea of fidelity and community [3].

## 4     About Open Science platform

Now consider the platforms and tools of one of the major European electronic infrastructures, EGI, which will cover how they can be the basis for an open science fund and then EOSC. EGI, an advanced computing engine for research, is a federated electronic infrastructure created to provide advanced computing services for research and innovation. EGI's infrastructure is primarily state-funded and has over 300 data centers and cloud providers throughout Europe and around the world. Its principles are based on an open academic community, and its mission is to create and provide open solutions for research and research infrastructures by combining digital capabilities, resources and expertise between communities and across national boundaries. EGI architecture is organized in platforms [3]:

― Basic Infrastructure Platform for Managed Distributed Infrastructure;
― Cloud infrastructure for managing the unified regional infrastructure;
― An open data platform that provides easy access to large and distributed data sets;
― A platform for cooperation, for the exchange of information and community co-ordination,
― Joint platforms, specialized service portfolios designed for specific academic communities.

The platform architecture allows any type and any number of shared platforms to coexist on physical infrastructure.

### 4.1     Augmented reality platform as a tool for open science

EGI launched the production phase of the cloud federation to serve research communities in May 2014, the EGI Federated Cloud. It integrates community, private and/or public clouds into a scalable computing platform for data and/or computing applications and services. Its architecture is based on the concept of an abstract cloud management environment (CMF), which supports a set of cloud interfaces for communities. Each Infrastructure Resource Center manages an instance of this CMF according to its own technological advantage and integrates it with the federation by interacting with the EGI's core infrastructure [3]. This integration is carried out using public interfaces supported by CMF, which minimizes the impact on the work of the site. Suppliers are organized in the area that uncover homogeneous interfaces and group resources dedicated to serving specific communities and/or platforms.

EGI Federated Cloud is based on a hybrid model where private, community, and public clouds can be integrated and already offer some tools that a service center must provide, such as virtualization and easy sharing and reuse of tools.

Each Infrastructure Resource Center manages an CMF instance according to its own technological advantage and integrates it with the federation by interacting with the EGI core infrastructure. Suppliers are organized in the areas of homogeneous interfaces (IaaS). Community platforms can use resources from one or more areas using these interfaces. AppDB VMOps enables the automatic deployment of virtual devices at all resource centers that support a specific community.

Olekcey O. Petrenko [9] explores the FIWARE directory as the main tool for creating web services for EOSC. Some of the services included in the FIWARE directory can be linked to the augmented reality:

− AEON Cloud Messaging: Real-time service provides cloud services (channels) for the transfer of unlimited number of entities, sharing unlimited amount of information, as well as services for managing actors involved in cloud environments.
− Complex Event Processing (CEP) – Proactive Technology Online: CEP analyzes real-time events by responding to situations rather than on individual events. Situations include composite events (for example, sequential), operator distribution by events (e.g., aggregation), and lack of operators.
− Cloud Rendering: The service defines a common way of requesting, receiving, and managing the video stream of a remote 3D application.

### 4.2  Practical application of augmented reality

Today, the Institute of Information Technologies and Training of the NAES of Ukraine is a partner of Visegrad Fund's Strategic Grant No. 21810100 "V4 + Academic Research Consortium for the Integration of Databases, Robotics and Language Technologies" [2]. As an example, let's look at one of the services developed by one of the partners of this project (Óbuda University Budapest, Magyarország), which can be included in the open science platform: MaxWhere (Hungary). MaxWhere combines several new technologies. The cognitive navigation technology (CogiNav) allows users to navigate smoothly across 3D spaces using only a laptop and mouse [5].

MaxWhere is the platform for managing all forms of digital content in 3D spaces. The main product – MaxWhere, which is largely similar to graphics engines (like Unity, Unreal), however, differs from them, since it has been optimized not for gaming applications, but for everyday digital life and professional industry. MaxWhere can be used in education and research.

Maxwhere [6] includes fast and innovative interfaces. This allows you to switch projects and go to different scientific communities, distribute research results in the fastest way. This is a combination of other applications that exist to organize the teamwork of scholars. 3D graphics will diversify your work without compromising performance. It can also be used by students to increase productivity and study data research.

Browser23 introduces a new web surfing philosophy: instead of having a limited number of tabs next to users, limiting their ability to switch between them and searching now, it allows you to set browser windows in 3D space, grouped by topics that are scaled for size and significance. The newly developed Ultra Sharing technology, which allows users to create VR offices that contain a large number of documents, and even complete the workflows of the project, and split these offices with one click [6]. Research shows that all these solutions combine an extremely effective way of visualizing, exchanging and manipulating large volumes of information while maintaining low cognitive load – a huge asset for understanding, configuring and managing large digital networking systems.

In 2017, MaxWhere was released as a tool for presenting 3D slides in interactive spaces. This solution is a blend between PowerPoint and Prezi, expanded with 3D objects. From a technological point of view, MaxWhere combines 3D space with web technologies. In this way, the world of open-source software (for example, Node.js, NPM and Node-RED) can be directed to MaxWhere applications.

## 5 Conclusions and prospects for further research

To date, the implementation of the European Research Area (ERA), as depicted by the European Council, can not be considered fully achieved. The implementation of an open and integrated environment for cross-border unimpeded access to advanced digital resources, services and opportunities facilitating the reuse of data and research services is accelerated by the initiative of the European Commission "European Open Science Cloud". Open science is seen as a natural paradigm for the promotion and development of such events. It can remove the barrier between neighboring communities, provide interdisciplinary cooperation, reinforce the need for knowledge sharing and allow free and unrestricted access. The advantages of the approach to open science and, in particular, the advantages of joint resources for the introduction of European infrastructure and the management of European open science were considered. We have analyzed the possible approach to the implementation of EOSC through open scientific communities. The EOSC architecture is based on the cloud hub federation, where the cloud hub provides data, services and features in a standard and consistent way. Hubs support the cloud provisioning paradigm to facilitate sharing, reuse, and combined data and tooling with virtualization. In addition, the federation of hubs provides a multi-layered organizational structure that complies with European policies, norms, restrictions and business models, and allows the creation of a community that can combine the various types of experiences available in each center. That is, an existing environment with several suppliers

EOSC is governed by special tools, processes and tools that determine the EOSC integration and management system owned, maintained, and developed by EOSC in accordance with the Commons management model. EOSC cloud nodes services are provided by many stakeholders: data providers, European research infrastructures, electronic infrastructures, research and local, regional and national institutions. The use of data directly benefits EOSC and the acceptance of open academic communities, using technologies, services and resources provided in the context of existing European electronic infrastructures. EOSC and electronic infrastructures can become a pole of engagement for designing and implementing appropriate solutions for managing and using a large number of data sets. This will allow you to create an integrated environment for rapid development, prototyping and service delivery for service platforms and scientific applications.